\documentclass{aa}
\usepackage{graphicx}
\begin{document}

\title{The diameter and evolutionary state of Procyon\,A}
\subtitle{Multi-technique modeling using asteroseismic and interferometric 
constraints}
\titlerunning{The diameter and evolutionary state of Procyon\,A}
\authorrunning{Kervella et al.}

\author{
P. Kervella\inst{1},
F.~Th\'evenin\inst{2},
P.~Morel\inst{2},
G.~Berthomieu\inst{2},
P.~Bord\'e\inst{3}
\and
J.~Provost \inst{2}}
\offprints{P. Kervella}

\institute{European Southern Observatory, Casilla 19001, Vitacura, Santiago 19, 
Chile
\and
D\'epartement Cassini, UMR CNRS 6529, Observatoire de la C\^ote 
   d'Azur, BP 4229, 06304 Nice Cedex 4, France
\and LESIA, Observatoire de Paris-Meudon, 5, place Jules Janssen, F-92195 Meudon 
Cedex, France}

\mail{Pierre.Kervella@eso.org}

\date{Received ; Accepted }

\abstract{
We report the angular diameter measurement obtained with the VINCI/VLTI 
instrument on the nearby star Procyon\,A ($\alpha$\,CMi\,A, F5IV-V),
at a relative precision of $\pm 0.9$\,\%. We obtain a uniform disk angular
diameter in the $K$ band of $\theta_{\rm UD} = 5.376\pm0.047$\,mas and a limb 
darkened 
value of $\theta_{\rm LD} = 5.448\pm0.053$\,mas.
Together with the {\sc hipparcos} parallax, this gives a linear
diameter of $2.048\pm0.025$\,D$_{\odot}$. We use this result in combination
with spectroscopic, photometric and asteroseismic constraints to model
this star with the CESAM code.
One set of modeling parameters that
reproduces the observations within their error bars are
an age of $2\,314$\,Myr, an initial helium mass fraction
$Y_{\rm i}=0.301$ and an initial mass ratio of heavy elements to
hydrogen $(\frac ZX)_{\rm i}=0.0314$.
We also computed the adiabatic oscillation spectrum of our model of
Procyon\,A, giving a mean large frequency separation of
$\overline{\Delta\nu_0}\approx 54.7\,\mu$Hz. This value is in
agreement with the seismic observations by Marti\'c et al.~(\cite{martic}, 
\cite{martic01}).
{ The interferometric diameter and the asteroseismic large frequency spacing
together suggest a mass closer to 1.4\,M$_{\odot}$ than to 1.5\,M$_{\odot}$}. 
We conclude that Procyon is currently ending its life on the main sequence,
as its luminosity class indicates.
}
\maketitle
\section{Introduction}
Procyon\,A ($\alpha$\,CMi, \object{HD 61421}, HIP 37279) is among the brightest 
stars
in the sky and is easily visible to the naked eye.
This made it an ideal target for a number of spectro-photometric calibration 
works.
It is also a visual binary star (ADS 6251A) classified F5 IV-V, with a white
dwarf (WD) companion orbiting the main component in 40 years.
The influence of this massive companion on the { apparent} motion of Procyon
was discovered by Bessel\,(\cite{bessel1844}).
Girard et al.\,(\cite{girard00}) have measured precisely the
orbit of the pair, and obtained masses of $1.497\pm0.037$~M$_{\odot}$
and $0.602\pm0.015$~M$_{\odot}$, respectively for Procyon\,A and B.
It has also been an asteroseismic target since a decade and
Martic et al.~(\cite{martic} and \cite{martic01} hereafter M99 and M01) have
measured a large frequency spacing of respectively 55 and 54~$\mu$Hz.
These asteroseismic observations provide strong constraints on stellar interior 
models, and on macroscopic stellar parameters like mass and radius. 
Comparing the direct interferometric measurements of this star to its model 
diameter is therefore important to cross-validate both approaches.

Matching Procyon's position in the HR diagram has been recognized as of great
difficulty by Guenther \& Demarque (\cite{gd93}, hereafter GD93). The reason is 
the poorly known mass and metallicity Z of the star. We note that among the computed  
models only their model $b$ is close to
the mean large frequency spacing measured by M99-M01.
Because numerous new studies and observational constraints (like the direct 
diameter) exist today,
we re-examine the status of Procyon\,A in helium content $\rm Y$ and in age.
We first review and present the adopted fundamental parameters of
Procyon\,A in Sect.\ref{param}, part of them being
used for the limb-darkening in Sect.~\ref{interf}, where we present
our new interferometric observations and the associated data processing.
In Sect.~\ref{modeling}, we detail several models of Procyon
computed with the CESAM code (Morel \cite{morel97}).
These models are constrained using the
spectroscopic effective temperature 
and the linear diameter value that we derived from the VINCI/VLTI observations. 
Thus, we avoid to fit the luminosity for which bolometric corrections are rather
uncertain for stellar luminosity class IV-V.
A study of the asteroseismic frequencies is then presented in  Sect.~\ref{astero},
before the conclusion.
\section{Fundamental parameters \label{param}}
Physical properties of Procyon\,A are well known thanks to the carefully 
measured orbit of the 
system and accurate {\sc hipparcos} parallax. Its mass has been recently 
measured by
Girard et al.\,(\cite{girard00}) at $\rm M_A=1.497\pm0.037$\,M$_{\odot}$,
adopting a parallax of $283.2\pm1.5$\,mas.
{ This mass is computed using a large number of observations of inhomogeneous 
quality.
The same authors
when they limit their observation sample to the excellent images obtained
with the WFPC2 found $\rm M_A = 1.465\pm0.041$\,M$_{\odot}$.
If we replace the parallax used by Girard et al.\,(\cite{girard00}) by the
parallax from the {\sc hipparcos} catalogue, $285.93\pm0.88$ mas, 
the sum of the masses of the binary decreases by 2.9\,\%. Consequently,
masses become respectively $\rm M_A=1.42\pm0.04$\,M$_{\odot}$
for Procyon\,A and $\rm M_B=0.575\pm0.017$\,M$_{\odot}$ for the white
dwarf Procyon\,B.
Allende Prieto et al. (\cite{allende02}, hereafter AP02), using the {\sc 
hipparcos} parallax of Procyon\,A, have estimated its mass to
be $1.42\pm0.06\,\rm M_{\odot}$.
From these works, we can conclude that the mass of
Procyon is probably between 1.4 and 1.5 $M_{\odot}$.
Consequently, we investigate these two values in our study.}
The other adopted stellar parameters are summarized in Table\,\ref{params}. 

The bolometric luminosity is given by Steffen (\cite{steffen}) 
$\log (L_{\star}/L_{\odot})=0.85 \pm 0.06$.
The photospheric properties of Procyon have been carefully 
studied by AP02, who derived an effective temperature 
$T_{\rm eff}=6512\,$K. This value is based on the angular diameter
measured in the visible by the Mark\,III optical interferometer
(Mozurkewich et al. \cite{mozurkewich91}).
Detailed 3D stellar atmosphere simulations have led AP02 to correct
this value of $T_{\rm eff}$ by 80\,K. Following their results, we adopt
$\rm T_{eff}=6530\pm 50$K in our computations with a surface gravity of $\log g 
= 3.96\pm0.02$.
It is interesting to notice that over the last twenty years, the effective 
temperature
estimates for Procyon are spread between 6545 and 6811\,K (Cayrel et 
al.\,\cite{cayrel}).
Adopting a color index of $\rm B-V=0.421$ and using the calibration by 
Alonso\,(\cite{alonso96})
we find  $T_{\rm eff}=6551\,$K which is very close to the value proposed by AP02.

The iron abundance at surface of [$\rm Fe/H]=-0.05\pm0.03$ dex with respect
to the solar one is also taken from AP02.
Therefore we adopt for the modeling of the internal structure of Procyon\,A a 
chemical
mixture which is calculated from the iron abundance using the approximation:
\begin{eqnarray}\label{eq:fesh}
\mathrm{[Fe/H]_{\rm s}} &\equiv&
\log\left(\frac{Z_{\rm Fe}}Z\right)_{\rm s}
+\log\left(\frac ZX\right)_{\rm s}
-\log\left(\frac{Z_{\rm Fe}}X\right)_\odot \\
&\simeq&\log\left(\frac ZX\right)_{\rm s}
 - \log\left(\frac ZX\right)_\odot \nonumber
\end{eqnarray}
where $\left(\frac{Z_{\rm Fe}}Z\right)_{\rm s}$ is the iron mass fraction within 
$Z$. We 
use the solar mixture of Grevesse \& Noels\,(\cite{gn93}) then
$\left(\frac ZX\right)_\odot=0.0245$. As we shall see in
Sect.~\ref{modeling}, we adopt at the age of Procyon\,A a surface abundance
$\left(\frac ZX\right)_{s}= 0.0217\pm0.0017$. Note that this adopted error bar 
of AP02
is very small.
Because the white dwarf (WD) Procyon\,B has already experienced
the AGB phase, it is not excluded that material coming from this
star can have contaminated the narrow surface convection layer of Procyon\,A.
However, there is no evidence of such an effect in the published table of 
abundances of Procyon\,A
(Steffen\,\cite{steffen}). Enriched elements like ``s'' or other elements coming 
from
Procyon\,B during its post-AGB phase do not seem to have migrated as gas or dust
in the atmosphere of Procyon\,A. { The mass transfer during this short period 
of the WD progenitor life could not exceed the accretion of the wind passing
close to the star.  Therefore, except if the two stars have filled in their Roche
lobe, the mass of Procyon\,A has only been changed by a negligible ammount.}

The projected rotational velocity of Procyon\,A has been determined by many 
authors who derived values smaller than 5.0\,km\,s$^{-1}$. Considering in particular
the value proposed by AP02, $v\sin i=3.2\pm0.5$\,km.s$^{-1}$,
we conclude that Procyon\,A is a slow rotator and we
neglect the rotational velocity in our modeling of its internal
structure. The semi-major axis of the Procyon orbit being
$\alpha \simeq 4\farcs 5$ (Girard et al.\,\cite{girard00}), the distance
between the two companions amounts to $\simeq 16 $\,{\sc au}.
The tidal interaction between the two components is therefore
negligible.
\section{Interferometric angular diameter \label{interf}}
\subsection{Instrumental setup and data processing}
The European Southern Observatory's Very Large Telescope Interferometer
(Glindemann et al.~\cite{glindemann}) is operated  on top of the Cerro Paranal, 
in Northern Chile,
since March 2001. The observations reported here were done with two test 
siderostats
(0.35 m aperture) and the VINCI beam combiner (Kervella et 
al.~\cite{kervella00}, \cite{kervella03a}). 
The Procyon visibility measurements were all obtained on the B3-D3 baseline, 24 
m in ground length.
We used a regular $K$ band filter ($\lambda=2.0-2.4\ \mu$m) for these 
observations.
The effective wavelength changes slightly depending on the spectral type of the 
observed target.
For Procyon, we estimate $\lambda_{\rm eff} = 2.178\pm0.003\ \mu$m. This error 
bar
adds quadratically a relative systematic uncertainty of 0.15\% on the final limb 
darkened
angular diameter error bar.

The raw data processing has been achieved using an improved version of
the standard data reduction pipeline (Kervella et al.~\cite{kervella03c}), whose 
general principle
is similar to the original FLUOR algorithm (Coud\'e du Foresto et al. 
\cite{cdf97}).
The two calibrated output interferograms are subtracted to remove residual
photometric fluctuations. Instead of the classical Fourier analysis, we
implemented a wavelet based time-frequency analysis (S\'egransan et al. 
\cite{s99}).
The output of the processing pipeline is a single value of the squared coherence 
factor
$\mu^2$ for each series of 500 interferograms and the bootstrapped error bar.

\subsection{Visibility calibration}
Sirius was chosen as the calibrator for Procyon observations as its limb 
darkened (LD)
angular diameter has been measured recently with high precision by
Kervella et al.~(\cite{kervella03d}), using VINCI and the 66m E0-G1 baseline.
Table~\ref{params} gives the relevant parameters of Procyon and Sirius.
The interferometric efficiency (IE) of the instrument was estimated from the 
$\mu^2$
values obtained on Sirius a short time before and after Procyon (typically 15 
minutes).
Thanks to the very high SNR of the Sirius interferograms, we were able to obtain 
very small
relative errors on the IE values, typically 0.2\% for a series of 500 
interferograms.
{
The error bar on the angular diameter of Sirius is treated as a systematic error
in the calibration process, and is therefore not averaged for multiple 
observations.
}
For the visibility fit, we took into account simultaneously the limb darkening 
and the
bandwidth smearing, as described in Kervella et al.~(\cite{kervella03b}, 
\cite{kervella03d}).

\begin{figure}[t]
\centering
\includegraphics[bb=0 0 360 288, width=8.5cm]{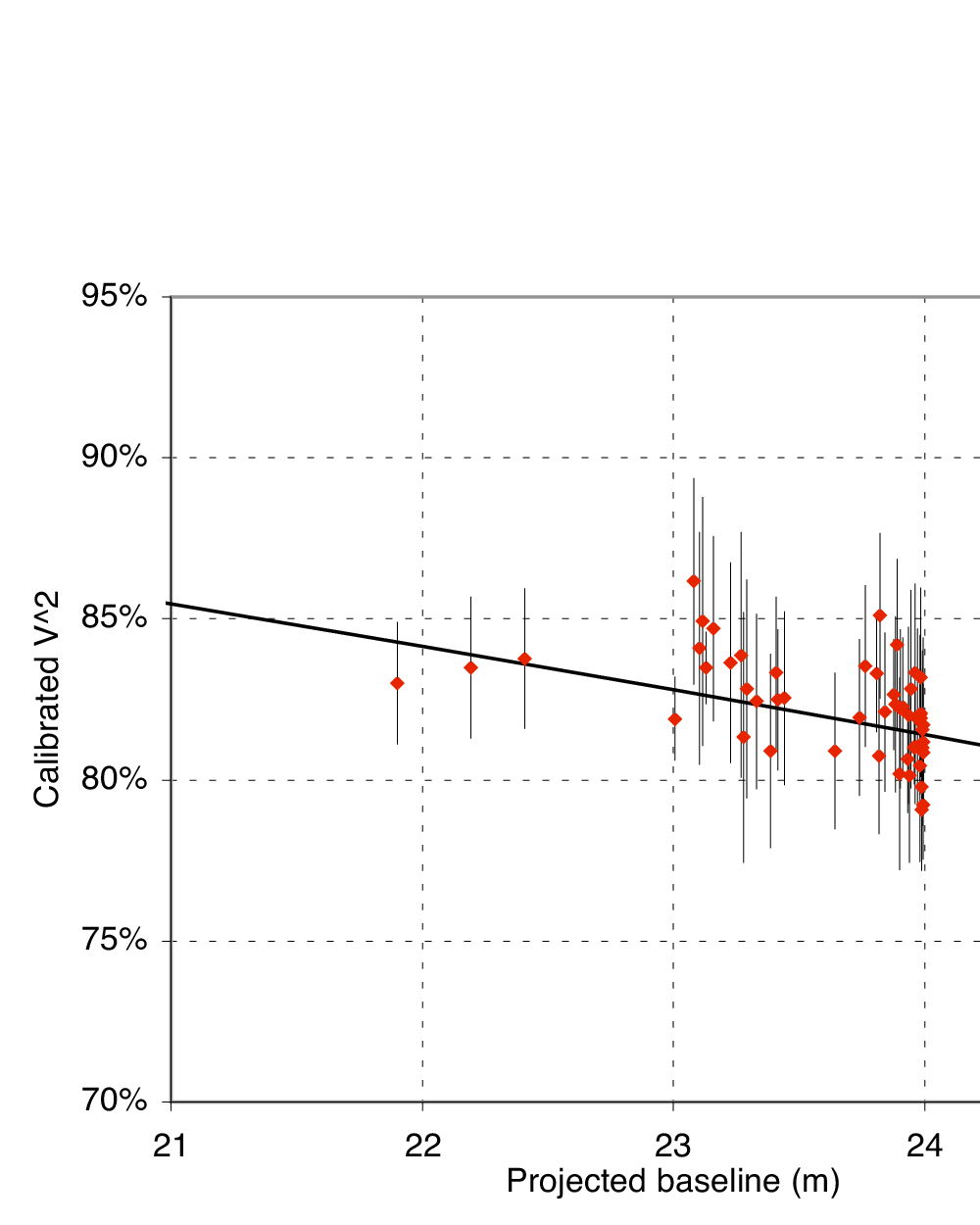}
\caption{Squared visibility measurements obtained on Procyon and best fit model 
(solid line).}
\label{visib_curve}
\end{figure}

\begin{figure}[t]
\centering
\includegraphics[bb=0 0 360 144, width=8.5cm]{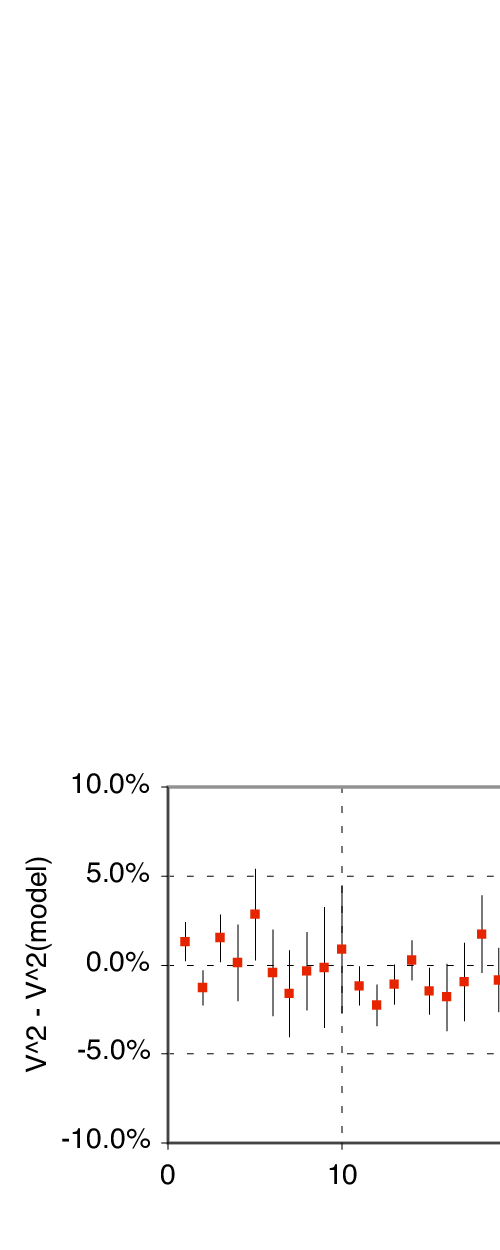}
\caption{Scattering of the Procyon measured visibilities around the best fit 
$V^2$ model. The plotted
error bars are the quadratic sum of the statistical and systematic 
contributions.}
\label{dispers_visib}
\end{figure}

Our February 2003 observation campaign of Procyon resulted in a total of 53 
calibrated $V^2$
measurements, from 23\,256 processed interferograms.
In the calibration process, we separated clearly the statistical and systematic 
error contributions.
This is essential to avoid underestimating the final error bars of the fit.
{
The latter corresponds to the uncertainty on the angular diameter of Sirius.
In spite of the fact that the angular size of Sirius is larger than Procyon's,
the small error on its measured value ($\pm$0.3\%) translates into a
small calibration error.
As a consequence, the global error on the angular size of Procyon
is dominated by the statistical contribution by a factor of two compared to the
systematic calibration error: the uncertainty on the angular size 
of
Sirius is not limiting our final precision.
}
The visibility points as a function of the projected baseline
are presented on Fig.~\ref{visib_curve}, and their residual scattering
around the best fit model on Fig.~\ref{dispers_visib}.
\begin{table}
\caption[]{Relevant parameters of $\alpha$\,CMi (Procyon) and its
calibrator $\alpha$\,CMa (Sirius). For Sirius, see also Kervella et al. 
(\cite{kervella03d}) 
\label{params}}
\begin{tabular}{lcc}
& $\alpha$\,CMi & $\alpha$\,CMa\\
& \object{HD 61421} & \object{HD 48915}\\
\hline
\noalign{\smallskip}
$m_\mathrm{V}$ & 0.34 & -1.47\\
$m_\mathrm{K}$ & -0.65 & -1.31\\
Spectral Type & F5IV-V & A1Vm \\
$M(M_{\odot})$ & $1.42\pm0.04$ & $2.12 \pm 0.02$\\
${\rm T_{\mathrm{eff}}}^{\mathrm{}}$ (K) & $6530\pm 50$ & $9900\pm200$\\
$\log g$ &  3.96 & 4.30\\
$[\mathrm{Fe}/\mathrm{H}]^{\mathrm{}}$ & -0.05$\pm0.03$ & 0.50$\pm0.20$ \\
$v\sin i$ & 3.16$\pm0.50$ & 16.0$\pm 1.0$ \\
$\pi^{\mathrm{a}}$ (mas) & $285.93\pm0.88$ & $379.22\pm1.58$\\
${\theta_{\rm UD}}$ (mas) & ${\bf 5.376\pm0.047}$ & $5.94\pm0.02^{\mathrm{b}}$\\
${\theta_{\rm LD}}$ (mas) & ${\bf 5.448\pm0.053}$ & $6.04\pm0.02^{\mathrm{b}}$\\
\noalign{\smallskip}
\hline
\end{tabular}
\begin{list}{}{}
\item[$^{\mathrm{a}}$] Parallax values from {\sc hipparcos} (Perryman et al. \cite{hip}).
\item[$^{\mathrm{b}}$] Sirius $\theta_{\rm UD}$ and $\theta_{\rm LD}$ are taken 
from Kervella et al.~(\cite{kervella03d}).
\end{list}
\end{table}

\subsection{Limb darkened angular diameter and linear size}
\begin{figure}[t]
\centering
\includegraphics[bb=0 0 360 144, width=8.5cm]{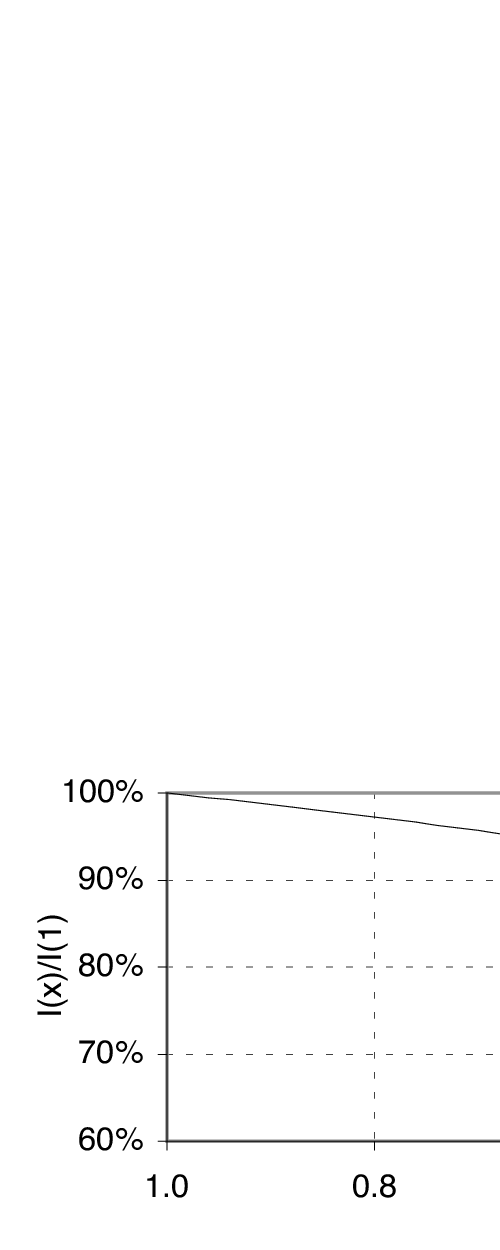}
\caption{Intensity profile of Procyon, from the four parameters law of 
Claret~(\cite{claret00}).}
\label{limb_darkening}
\end{figure}

%
The intensity profile $I(\mu)/I(1)$ that we chose for Procyon was
computed by Claret et al.~(\cite{claret00}), based on the ATLAS9 model
atmospheres (Kurucz\,\cite{kurucz92}). It is a four parameters law:
%
\begin{equation}
{I(x)}/{I(1)} = 1 - \sum_{k=1}^{4}{a_k(1-x^{\frac{k}{2}}})
\end{equation}
which coefficients are $a_1=0.5089$, $a_2=0.0822$, $a_3=-0.3978$, and 
$a_4=0.1967$. The parameter $x=\cos(\theta)$ is the cosine of the
azimuth of a surface element of the star.
The resulting profile is shown on Fig.~\ref{limb_darkening}.
AP02 have identified an influence of the convective granulation of
Procyon on its limb darkening at visible wavelengths. Though, the amplitude of 
this
effect is already very small at $\lambda = 1.0\,\mu$m, and we neglect it for our
observations at $\lambda = 2.2\,\mu$m.

We obtain directly the LD angular diameter of Procyon from a classical $\chi^2$ 
minimization,
$\theta_{\rm LD} = 5.448\pm0.053$~mas.
Using a simple uniform disk model, we find a value of
$\theta_{\rm UD} = 5.376\pm0.047$~mas.

We can compare the LD value with the previously published interferometric
measurements, listed in Table~\ref{published_diams}.
Our value of $\theta_{\rm LD}$ is compatible with previous measurements,
that were all obtained using visible wavelength observations.
It is also in excellent agreement with the average of all published values.
From the VINCI/VLTI value of $\theta_{\rm LD}$ and the {\sc hipparcos} parallax, 
we deduce a
linear diameter of $2.048\pm0.025$ D$_{\odot}$. Computing the average of all 
published $\theta_{\rm LD}$ values, we obtain a linear radius of
$\rm R_{\star}=2.047\pm0.020$ R$_{\odot}$ for Procyon~A,
statistically identical to our result. 

\begin{table}

\caption[]{Interferometric measurements of the angular diameter of Procyon\,A
from the literature.\label{published_diams}}
\begin{tabular}{lccc}
Instrument & $\lambda$ ($\mu$m) & $\theta_{\rm UD}$ (mas) & $\theta_{\rm LD}$ 
(mas) \\
\hline
\noalign{\smallskip}
Intensity interf.$^{\mathrm{a}}$ & 0.45 & $5.10\pm0.16$ & $5.41\pm0.17$ \\
Mark III$^{\mathrm{b}}$ & 0.80 & $5.26\pm0.05$ &  \\
Mark III$^{\mathrm{b}}$ & 0.45 & $5.14\pm0.05$ &  \\
Mark III$^{\mathrm{c}}$ & 0.80 & & $5.46\pm0.08$ \\
NPOI$^{\mathrm{c}}$ & 0.74 & $5.19\pm0.04$ & $5.43\pm0.07$ \\
VINCI/VLTI & 2.18 & ${\bf 5.38\pm0.05}$ & ${\bf 5.45\pm0.05}$ \\
\hline
\noalign{\smallskip}
Average value & & & $5.445\pm0.035$ \\
\noalign{\smallskip}
\hline
\end{tabular}
\begin{list}{}{}
\item[$^{\mathrm{a}}$] Hanbury Brown et al.~(\cite{hanbury74}).
\item[$^{\mathrm{b}}$] Mozurkewich et al.~(\cite{mozurkewich91}).
\item[$^{\mathrm{c}}$] Nordgren et al.~(\cite{nordgren01}).
\end{list}
\end{table}

\section{Modeling \label{modeling}}
\begin{figure}[t]
\centering
\includegraphics[width=6.cm, angle=-90]{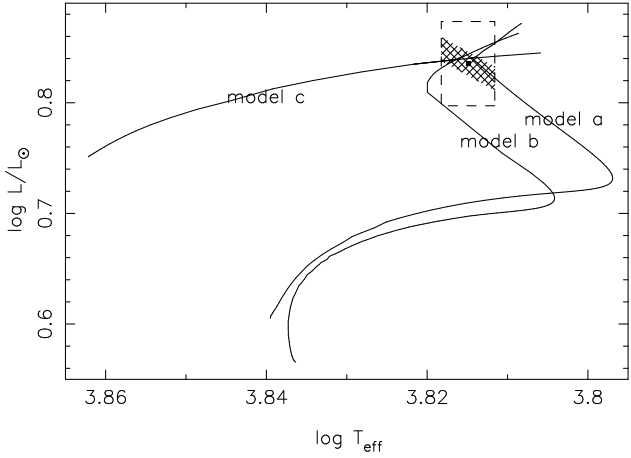}
\caption{Evolutionary tracks in the HR diagram of models of Procyon\,A (see 
Table~\ref{tab:mod}).
Dashed rectangle delimits the uncertainty domain for luminosity and effective 
temperature,
while the hatched area delimits the uncertainty domain for effective temperature
and the interferometric radius.
\label{fig:HR}}
\end{figure}

The modeling of Procyon\,A is based on the matching of 
spectroscopic effective temperature, surface metallicity
and interferometric radius by { a ``satisfactory'' evolutionary model for a given mass.}
The constraint between $R_{\star}$ and $T_{\rm eff}$ within their error bars is
illustrated in Fig.~\ref{fig:HR} by the hatched parallelogram, while the
constraint between the $T_{\rm eff}$ and $L_{\star}$ is
illustrated by the dashed rectangle.
That emphazises the advantage of the use of the measured $R_{\star}$ instead
of the $L_{\star}$ which depends of photometric calibrations and bolometric
corrections.

Table~\ref{tab:mod} shows the characteristic and initial modeling parameters
of the star.
The ordinary assumptions of stellar modeling are made, i.e. spherical symmetry,
no rotation, no magnetic field and no mass loss. The relevant nuclear reaction 
rates
are taken from the NACRE compilation by Angulo et al.\,(\cite{angulo99}). The 
equation of
state adopted is taken from Eggleton et al.\,(\cite{eff73}), and
the opacities are from the OPAL database (Iglesias \& 
Rogers\,\cite{iglesias96}), using the
Grevesse \& Noels\,(\cite{gn93}) mixture.
The microscopic diffusion is described using the formalism of 
Burgers\,(\cite{burgers69}) with the resistance coefficients of Paquette et 
al.\,(\cite{pa86}).
We take into account the radiative diffusivity as
recommended by Morel \& Th\'evenin\,(\cite{morel02}), that limits the efficiency
of the microscopic diffusion in outer-layers of stars with intermediate masses.
The atmosphere is restored using Hopf's law (Mihalas\,\cite{mi1978}).
The definition of the radius of the star is the bolometric radius, 
where $T(R_{\star}) = T_{\rm eff}$.
In the convection zones the temperature gradient is
computed according to the $\rm MLT_{CM}$ convection
theory with a mixing length parameter of $\Lambda=1$
(Canuto \& Mazzitelli \cite{cm91}, \cite{cm92}).

\begin{figure}[t]
\centering
\includegraphics[width=6.cm, angle=-90]{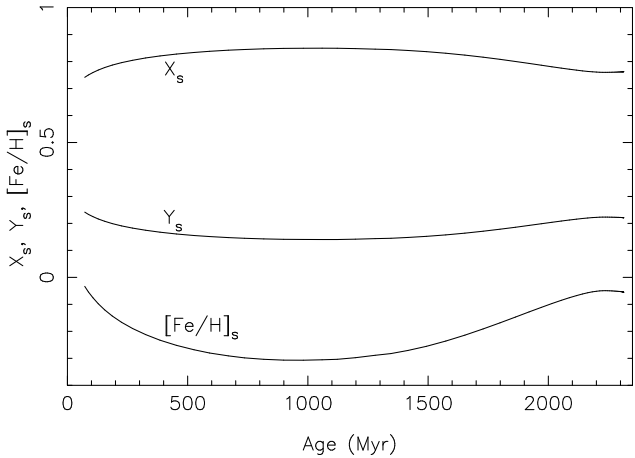}
\caption{Changes with respect to time for $X_{\rm s}$, $Y_{\rm s}$
and [Fe/H]$_{\rm s}$ for model $a$ of Procyon.
\label{fig:XYZ}}
\end{figure}

{
Following the discussion of Sect.~\ref{param} we investigate the sensitivity
of our models to a variation of the mass of Procyon in the range 1.4 to 1.5 $M_{\odot}$.}
We obtained a ``satisfactory'' model $a$ (Fig.~\ref{fig:HR}), detailed in 
Table~\ref{tab:mod}, which reaches the hatched parallelogram
corresponding to an age of 2\,314 Myr. 
Its evolutionary state corresponds closely to the end of the main sequence,
when the convective core has disappeared owing to the exhaustion of the hydrogen 
at center.
We present on Fig.~\ref{fig:XYZ} the variation for model $a$ of $ X_{\rm s}, Y_
{\rm s}$ and of
$\rm [Fe/H]_s$ as a function of the age. 
From the ZAMS until about 1 Gyr, due to the gravitational settling,
the surface abundances of helium and heavy elements ({\em resp.} hydrogen)
decrease ({\em resp.} increases).
After 1 Gyr, the density in the envelope decreases
due to the enlargement of radius. This leads to an increase of
the radiative mixing, with the consequence of a dredge-up which increases
({\em resp.} decreases) $Y_{\rm s}$ and $\rm [Fe/H]_s$ ({\em resp.} 
$ X_{\rm s}$) at the surface.

We also computed a ``satisfactory'' model $b$ without microscopic diffusion in order to
estimate its effect on the age of the star and we found that suppressing
the diffusion increases the age by $\sim$ 400 Myr. The evolutionary state of 
this model is the beginning of H burning in shell.

Following the prescriptions of Schaller et al.\,(\cite{schaller92})
we have computed additional models that include overshooting of the
convective core of radius $R_{\rm co}$ over the distance
$O_{\rm v}=A\,\min(H_{\rm p},R_{\rm co})$. 
It has been impossible to find a ``satisfactory'' model reaching
the narrow hatched area of the HR diagram with parameters $A>0.03$.

Finally, we computed a model $c$ with Girard et al.'s (\cite{girard00})
mass : 1.50 $M_{\odot}$ and find an age of 1.3 Gyr without overshoot.
We also computed a model with an overshoot of 0.15 in agreement with Ribas, 
Jordi \& Carme\,(\cite{ribas}) for stars of mass lower than 1.5\,M$_{\odot}$ 
resulting in a model which increases the age of the star to 1.55 Gyr.

The parameters for all models are listed in Table~\ref{tab:mod}.
Fig.~\ref{fig:HR} shows the corresponding evolutionary tracks in the {\sc hr} 
diagram.
We note that our models fit the value of $\log g=3.96$ proposed by AP02.
\begin{table}
\caption[]{
Procyon\,A  models (without overshoot, see text) 
lying within uncertainty box in the {\sc hr} diagram.
The subscripts ``$_{\rm i}$'' and ``$_{\rm s}$'' respectively refer
to initial values and surface quantity at present day. ``$_{\rm c}$''
refers to the central value.
}\label{tab:mod}
\begin{tabular}{lcccc} \\
\hline
Model                         &     a     &     b     &     c     \\
\hline
\noalign{\smallskip}

$m/M_\odot$                       &   1.42    &   1.42    &   1.50 \\
$Y_{\rm i}$                       & 0.3012 &    0.2580 & 0.345    \\
$Y_{\rm s}$                       & 0.2209 &    0.2580 & 0.202    \\
$\left(\frac ZX\right)_{\rm i}$   & 0.03140 &   0.0218 & 0.0450    \\
$\left(\frac ZX\right)_{\rm s}$   & 0.02157 &   0.0218 & 0.0220   \\
diffusion                         & yes     &   no     & yes \\
$X_{\rm c} $                     & 0.00051 &   0.00000 & 0.2180    \\
age (Myr)                         & 2\,314 &    2\,710 & 1\,300   \\
$T_{\rm eff}$\,(K)                & 6524 &      6547 & 6553    \\
$\log g$                          & 3.960 &     3.967 & 3.994    \\
$\rm [Fe/H]_{\rm i}$              & +0.107 &    -0.051 & +0.264    \\
$\rm [Fe/H]_{\rm s}$              & -0.055 &    -0.051 & -0.043    \\
$\log(L/L_\odot)$                       & 0.8409 &    0.8405 & 0.8390    \\
$ R/R_\odot$                      & 2.0649 &     2.0495 & 2.0420 \\

$\overline{\Delta\nu_0}\,(\mu$Hz) & 54.7 &    55.4  & 56.4   \\

\noalign{\smallskip}
\hline
\end{tabular}
\end{table}

\section{Asteroseismology\label{astero}}
{ 
Asteroseismic  observations of Procyon\,A have been reported by
M99 and M01 and indicate an excess of power around 1 milliHertz with
a probable frequency spacing of 54 to 55 $\mu$Hz.

 The narrow convective envelope of the star we consider
may stochastically excite solar-like oscillations. As in the Sun, these 
oscillations will have quasi equidistant frequencies at given degree.

For all the models, we have computed a
set of adiabatic frequencies of the normal modes 
for degrees $\ell=0,1,2,3$. The set of frequencies of model  $a$
are given in the frequency range $450$ to $1\,600\,\mu$Hz
in Table~\ref{tab:frequence}.

The large frequency spacing is defined as the difference between
frequencies of modes with consecutive radial order $n$:
$\Delta\nu_\ell(n) \equiv \nu_{n, \ell} - \nu_{n-1,\ell}.$
In the high frequency range, i.e. large radial orders,
$\Delta\nu_\ell$ is almost constant with
a mean value $\overline{\Delta\nu_0}$, strongly related
to the mean density of the star. The computed frequencies are
fitted to the following asymptotic relation (Berthomieu et al. \cite{ber93}):
$\nu_{n,\ell} = \nu_0 +\overline{\Delta\nu_\ell} (n-n_0) +a_\ell(n-n_0)^2$.
With $n_0= 21$,   $\ell = 0$ and radial order $n$ between 17 and 26
we obtain an  estimate of the mean large  difference $\overline{\Delta\nu_0}$.

The mean large spacing of model $a$ 
is 54.7 $\mu$Hz which fits well observational predictions of M99 \& M01
($\overline{\Delta\nu_0} \sim 54-55\,\mu$Hz). 
{
For the model $c$, the mean large spacing is found to be 56.4 $\mu$Hz, 
significantly different from the observed value.
A mass as large as 1.5\,$M_{\odot}$ for Procyon~A seems therefore improbable.
Adopting the VINCI/VLTI radius, and looking at the scaling of Kjeldsen \& 
Bedding~(\cite{kb95}),

\begin{equation}\label{eq_sismo1}
$$\overline{\Delta\nu_{\rm osc}}\sim 134.9\sqrt{\frac 
{m/M_\odot}{({R_\star/R_\odot})^3}}\ (\mu Hz),$$
\end{equation}
the most straightforward way to decrease $\overline{\Delta\nu_0}$
is to decrease the mass of  Procyon~A.

Because the luminosity L is in relation with $T_{\rm eff}$ and with the radius $R$
through the formula $L/L_{\odot}= (R/R_{\odot})^2 (T_{\rm eff}/T_{\rm eff}^{\odot})^4$
we find from Eq.~\ref{eq_sismo1} the expression of the mean large spacing
as a function of $L$ and $T_{\rm eff}$,
\begin{equation}\label{eq_sismo2}
$$\overline{\Delta\nu_{\rm osc}}\sim 134.9{\frac 
{\sqrt{m/M_\odot}}{(L/L_{\odot})^{3/4}(T/T_{eff})^3}}\ (\mu Hz). $$
\end{equation}
From this formula it is easy, if needed, to estimate the derivatives of the mean large spacing
with respect to the mass, the effective temperature and the
luminosity. To estimate the dependance between $\overline{\Delta\nu_{\rm osc}}$
and $Z_i$ we have computed a modified model $a$ with $\Delta Z_i=0.001$ and found
a variation in the parallelogram error box of : $\Delta T_{eff}=22 K$, $\Delta L/L=0.008$,
and $\Delta_{age} = 33$Myr.
There were no significant change with $Z_i$ of $\overline{\Delta\nu_{\rm osc}}$
as can be expected from Eq.~\ref{eq_sismo1}.
}}

\begin{table}
\caption[]{Asteroseismic frequencies ($\mu$Hz) of Procyon\,A  for model $a$.
The ``$\star$'' correspond to ``$g$'' modes which are presents in the frequency
domain due to the evolutionary state of the star.
}
\label{tab:frequence}
\begin{tabular}{ccccc} \\
\hline
   $\ell$=0     &$\ell$=1    & $\ell$=2    & $\ell$=3    &   n  \\
\hline
\noalign{\smallskip}
   485.35 &  513.36 &  532.31&   551.46  &  8\\
   --     &  --      & --    &  597.20 &   $\star$\\
   537.93 &  563.43  & 585.90  & 605.61&    9\\
   591.09 &  615.78 &  638.60   &657.72   &10\\
   643.23 &  666.79&   689.80&   708.98  & 11\\
   694.17 &  717.44   &741.00 &  760.87 &  12\\
   745.32 &  769.15  & 793.75  & 814.22&   13\\
   798.11 &  822.24 &  847.62   &868.87   &14\\
   --     &  --      &896.38&    --     &  $\star$  \\ 
   852.14 &  876.76  & 904.48 &  923.72 &  15 \\
   906.98 &  931.22 &  957.38  & 978.12&   16 \\
   961.42 &  985.26   &1011.19  &1031.92   &17\\
   --     &  --     &   --  &  1075.73  &  $\star$ \\
  1015.27 &  1038.85&  1065.03 & 1086.34 &  18\\
  1069.16 &  1092.64  &1119.00  &1140.31&   19\\
  1123.16 &  1146.76 & 1173.64&  1195.27   &20\\
  1177.86 &  1201.57&  1228.79 & 1250.68  & 21\\
  1233.05 &  1256.82  &1284.20  &1306.18 &  22\\
  1288.46 &  1312.09 & 1339.65&  1361.81&   23\\
  1343.91 &  1367.55&  1395.21 & 1417.43   &24\\
  1399.42 &  1422.95  &1450.65  &1472.94  & 25\\
  1454.83 &  1478.24 & 1506.00&  1528.39 &  26\\
  1510.12 &  1533.53&  1561.40 & 1583.89&   27\\
\noalign{\smallskip}
\hline
\end{tabular}
\end{table}

\section{Discussion \& Conclusions}
We have reported in this paper our modelisation of Procyon\,A based on
observations by long-baseline interferometry and asteroseismology.
The use of the measured stellar diameter allows to reduce significantly the 
error bar on the luminosity of the star, as it does not require any bolometric
corrections. This advantage has been demonstrated in this study and in the modeling of 
Sirius by Kervella et al.\,(\cite{kervella03d}).

Our model shows that using the given set of parameters and given physics,
Procyon A is currently finishing to burn its central hydrogen, and is at the 
phase where
the convective core is disappearing.
The error on the measured radius gives a narrow uncertainty of 10 Myr
on the deduced age. Provencal et al. (\cite{pro02}) have discussed the cooling 
time
of the WD Procyon\,B and found that the progenitor
ended its lifetime $1.7\pm0.1$\,Gyr ago. 
We derive an age of 2\,314 Myr for Procyon\,A.
Subtracting the cooling age of the WD companion to our determination of the
age of Procyon\,A leads to a lifetime of 614 Myr for the progenitor of 
Procyon\,B.
This indicates that the mass of the progenitor is approximately 
2.5\,M$_{\odot}$.
This value yields a mass of $\sim 0.57M_{\odot}$ for the core of the
corresponding Thermal-Pulsating-AGB star (for Z=0.020)
(Bressan et al. \cite{bre93}), which is the minimum
possible value for the final mass of the WD (see also Jeffries \cite{jeffries}).
{
This estimate of 0.57\,$M_{\odot}$ agrees very well with the mass of
Procyon\,B that we deduced from Girard et al.\,(\cite{girard00}).
We note that the age obtained with model $c$ is shorter than the cooling age of
Procyon~B. This argument suggests a mass lower than 1.5 M$_{\odot}$
for Procyon~A and strengthens the asteroseismology results.

Further progress on the modeling of Procyon will be possible
when the accuracy on the flux of the star is improved to less than 1\,\%.
Waiting for such accuracy, the uniqueness of the solutions resulting from
computed models fitting a narrow box in the HR diagram will come from future
detailed asteroseismic studies. For example, other linear combination of 
frequencies like the small frequency spacings
(see e.g. Gough \cite{go91}) will constrain the age and the mass of the star.
} We therefore encourage observers to progress on the asteroseismology
of this star.

Large uncertainties also come from the adopted chemical 
abundance mixture $Z_s$ which is still rather uncertain.
Only a few chemical element abundances 
are measured today and most of them with a low accuracy. This uncertainty 
on $Z_s$ is the largest source of error on the estimated initial helium content
$Y_i$ and on the age of Procyon.
Thus, we recommend that surface abundances should be derived
from 3D atmosphere studies, in particular for oxygen and other important
donors of electrons.

\begin{acknowledgements}
The VLTI is operated by the European Southern Observatory at Cerro Paranal, 
Chile.
The VINCI public commissioning data reported in this paper
has been retrieved from the ESO/ST-ECF Archive.
The VINCI pipeline includes the wavelets processing technique,
developed by D. S\'egransan (Obs. de Gen\`eve).
No VLTI observation would have been possible without the efforts of the
ESO VLTI team, for which we are grateful.
This research has made use of the SIMBAD database at CDS, Strasbourg (France).
\end{acknowledgements}

{}
\end{document}